Check for updates

OPEN

DATA DESCRIPTOR

# Future Deployment and Flexibility of Distributed Energy Resources in the Distribution Grids of Switzerland

Lorenzo Zapparoli[1,3], Alfredo Oneto[1,3], María Parajeles Herrera[2], Blazhe Gjorgiev[1], Gabriela Hug[2] & Giovanni Sansavini[1] ✉

**The decarbonization goals worldwide drive the energy transition of power distribution grids, which operate under increasingly volatile conditions and closer to their technical limits. In this context, localized operational data with high temporal and spatial resolution is essential for their effective planning and regulation. Nevertheless, information on grid-connected distributed energy resources, such as electric vehicles, photovoltaic systems, and heat pumps, is often fragmented, inconsistent, and unavailable. This work introduces a comprehensive database of distributed energy resources and non-controllable loads allocated in Switzerland's medium- and low-voltage distribution grid models, covering over 2 million points of connection. Remarkably, this data specifies the flexibility capabilities of the controllable devices, with a set of projections aligned with national forecasts for 2030, 2040, and 2050. The database supports studies on flexibility provision of distributed energy resources, distribution grid resilience, and national energy policy, among other topics. Importantly, its modular structure allows users to extract national- and local-scale information across medium- and low-voltage systems, enabling broad applicability across locations.**

## Background & Summary

Power systems consist of transmission, sub-transmission, and distribution grids, with the latter further divided into medium- and low-voltage systems[1]. Transmission grids transport hundreds or even thousands of megawatts over long distances from large generation stations, where step-up transformers increase the voltage for efficient power transfer. Sub-transmission systems interconnect transmission and distribution grids and segment the power system. Ultimately, distribution grids supply residential, commercial, and industrial loads, which are evolving fast due to the massive deployment of distributed energy resources (DERs)[2]. The adoption of these resources is also expected to rise due to the ongoing global energy transition to low-carbon systems. DERs include rooftop solar photovoltaic (PV) systems, battery energy storage systems (BESSs), heat pumps (HPs), and electric vehicles (EVs). Their integration can, on the one hand, lead to overvoltage, reverse power flows, and overloading components; on the other hand, their control can provide operational flexibility[3]. Consequently, the implication of DERs' penetration raises pressing questions about operation (e.g., reactive power provision from inverters[4,5]), system planning (e.g., needs for line expansions[6]), and energy policy (e.g., incentive programs[7,8]).

High spatial resolution data on DER deployment and operation are essential for addressing current energy challenges. However, such data remain scarce due to privacy concerns and the limited digitalization of distribution networks. Some research efforts have aimed to fill this gap. For instance, non-intrusive load monitoring methods have helped identify DER profiles from aggregated measurements on a distribution transformer's low-voltage side[9]. Stochastic models have been developed to capture spatially resolved intermittent load fluctuations[10]. In addition, another study presents a dataset of DERs integrated into a U.S. test transmission grid, reflecting current adoption levels of PVs, BESSs, and EVs[11]. Despite the usefulness of these datasets, no previous work has incorporated the deployment of DERs in geo-referenced distribution grids across large-scale areas. Developing such a database would enable testing operational strategies and support policy-relevant analyses for the region under study.

[1]Reliability and Risk Engineering Laboratory, Institute of Energy and Process Engineering, Department of Mechanical and Process Engineering, ETH Zürich, Zürich, Switzerland. [2]Power Systems Laboratory, Institute for Power Systems & High Voltage Technology, Department of Information Technology and Electrical Engineering, ETH Zürich, Zürich, Switzerland. [3]These authors contributed equally: Lorenzo Zapparoli, Alfredo Oneto. ✉e-mail: sansavig@ethz.ch

This article presents a dataset that includes geo-referenced allocations and high temporal-resolution time series for DERs and non-controllable loads in Switzerland's distribution grids[12]. These synthetic grid models cover the country, comprising medium- and low-voltage levels. Scenarios for DER penetration are projected for 2030, 2040, and 2050, consistent with national energy strategy targets. Moreover, year-round time profiles with hourly resolution are provided for: 1) expected PV power generation and its standard deviation, 2) outdoor temperature (used to model HP demand), 3) EV base charging and lower/upper shifting limits, and 4) non-controllable load profiles. In addition to time series and devices' spatial locations, the dataset characterizes the operational flexibility of each DER class: PV systems through curtailment, BESSs via controllable state of energy (SOE), HPs through the thermal inertia of buildings and indoor temperature flexibility, and EVs through smart charge shifting (the so-called V1G). In total, the dataset covers more than 2 million connection points (nodes) across the Swiss medium- and low-voltage grids.

The dataset's compatibility with existing Swiss grid models and DER penetration scenarios makes it suitable for various studies. Parts of this dataset have already informed previous work by the authors and fostered interest in developing a country-wide, high-resolution version. For instance, one study quantified maximum grid loadability using information on PV and medium-voltage consumption profiles[13]. Another study used data for a Swiss low-voltage grid to design new ancillary services products for DERs[14]. Furthermore, this dataset will support additional studies, given its modular structure and spatiotemporal resolution. Its features enable researchers to analyze diverse locations by extracting national and local-scale information across medium- and low-voltage systems.

Potential future research applications of the dataset include, but are not limited to:

- Reliability assessment of distribution grids[15], considering the effects of spatially heterogeneous DER deployment.
- Identification of feasible operating regions[16,17] to characterize the allowable range of nodal power injections and demand, and derivation of power feed-in limitations to support planning[18].
- Machine learning applications, including training deep learning solvers for distribution power flow[19] and clustering of graph-structured distribution data[20].
- Operational studies addressing privacy-preserving optimal power flow[21] and DER control via distributed optimization[22].
- Grid expansion planning research considering multistage investments[23] and DER coordination to manage peak load[24].

Researchers can also apply the methods developed in this work to generate datasets of DER deployment and flexibility modeling in other locations. These methods may support not only the analysis of existing distribution systems, but also the assessment of regional electrification using geo-spatial data[25,26].

## Methods

The input datasets are summarized in Table 1. They support the derivation of DER deployment, time series for power production and consumption, and flexibility capabilities for 2030, 2040, and 2050. This section describes the processing steps for each DER category and non-controllable loads, their integration into the distribution grids, and their flexibility capabilities. Figure 1 provides an overview of the dataset construction.

### Photovoltaic data.

The PV data source provides geospatial information on rooftop PV potential across Switzerland[27]. The rooftops include virtual PV system installations, based on geometric projections to tilted roofs and south-facing rows at a 30° tilt for flat roofs. Although orientation and tilt influence generation yield, their selection depends on various objectives and constraints, including dynamic greenhouse gas emissions and grid export limitations[28]. Modifications to PV angles are therefore omitted. Moreover, while relevant to Switzerland's energy transition, ground-mounted PV is not included in the DER dataset. Its deployment is considered a planning decision, and a substantial share, mostly Alpine PV, is expected to connect at the transmission level, which falls outside the dataset's scope.

The input data supplies yearly PV generation per rooftop using twelve representative days, one for each month. It includes daily average generation profiles, the corresponding standard deviation, and the monthly average of global tilted irradiance for each roof. Since the nominal installed PV capacity per building is not reported, it is computed using PV panel type specifications[27]. Starting from the PV panel area ($A^{\text{panel}}$), its rated power ($P^{\text{rated}}$), and the PV system efficiency ($\eta$), the nominal installed capacity per building is computed as:

$$\overline{P_i^{PV}} = \frac{1}{H}\sum_{j=1}^{H}\left(\frac{P_{i,j}^{out}}{\eta \cdot G_{i,j} \cdot A^{panel}}\right)P^{rated} \quad \forall i \in \mathcal{B}, \tag{1}$$

where the sum considers all hours of the year ($H = 8760$), $\mathcal{B}$ is the set of buildings (roofs) in the PV dataset, and $P_{i,j}^{out}$ and $G_{i,j}$ represent the power output and tilted solar irradiance at building rooftop $i$ during hour $j$, respectively.

The PV allocations for 2030, 2040, and 2050 are generated based on national penetration projection values from the Swiss Federal Office of Energy (SFOE)[29], as shown in Table 2. A fraction of all buildings with rooftop PV potential is uniformly randomly selected to meet the PV deployment targets. This procedure ensures consistency with the projections while preserving the spatial and temporal characteristics of the original data. Furthermore, to preserve continuity in PV deployment across years, all buildings with PV installations in a given year retain their systems in subsequent years.

| Dataset | Description |
|---|---|
| PV (*) | **Duration**: One year. |
| | **Resolution**: One hour. |
| | **Source**: Scientific paper and its dataset[27,52] (https://doi.org/10.5281/zenodo.3609833). |
| | **Comment**: PV generation potential for rooftop PV at the national level in Switzerland with single-building resolution. Each building reports the hourly-resolved yearly production profile, standard deviation of production, and global tilted irradiance. Data is provided in the form of one representative day per month. |
| Temperature (△) | **Duration**: One year. |
| | **Resolution**: One hour. |
| | **Source**: Web database[53] (https://www.meteoswiss.admin.ch/services-and-publications/service/open-data.html). |
| | **Comment**: Ambient temperature measurements for weather stations in Switzerland. Data is provided as a temperature profile with an hourly resolution for each weather station. |
| EV (*) | **Duration**: One year. |
| | **Resolution**: One hour. |
| | **Source**: Scientific paper and its dataset[41,42] (https://doi.org/10.5281/zenodo.15194722). |
| | **Comment**: Switzerland's EV consumption and flexibility potential at the level of municipalities for the entire country. |
| Low-voltage non-controllable load (*) | **Duration**: One year. |
| | **Resolution**: One hour. |
| | **Source**: Scientific paper and its dataset[43] (https://data.sccer-jasm.ch/demand-hourly-profile-retrofits-cesar/latest/). |
| | **Comment**: Hourly-resolved yearly electricity demand profiles for a selected set of archetype buildings in Switzerland. It differentiates between single-family houses, multi-family houses, offices, schools, shops, restaurants, and hospitals. |
| Medium-voltage non-controllable load (△) | **Duration**: Ten years. |
| | **Resolution**: Fifteen minutes. |
| | **Source**: Web database[45] (https://data.stadt-zuerich.ch/dataset/ewz_stromabgabe_netzebenen_stadt_zuerich). |
| | **Comment**: The 15-minute resolved profile of the electricity demand of the city of Zurich for different voltage levels. Data are available from 2015 until 2025. |
| Building registry (△) | **Source**: Web database[34] (https://www.housing-stat.ch/de/madd/public.html). |
| | **Comment**: Information on buildings in Switzerland, including unique identifiers, geographic data (coordinates, address), and structural characteristics such as dimensions, number of floors, and volume. It includes the building's status and construction/demolition periods. Key technical installations, including heating and hot water systems, are also reported. |
| Building archetypes (*) | **Source**: Scientific paper and its dataset[35] (https://www.mdpi.com/article/10.3390/buildings13010040/s1). |
| | **Comment**: Set of archetype buildings for Switzerland indexed to the construction year and construction type. For each archetype typical thermal properties, such as thermal transmittance, are reported. |
| Municipality geo-information (△) | **Source**: Web database[51] (https://www.swisstopo.admin.ch/de/landschaftsmodell-swissboundaries3d). |
| | **Comment**: Municipalities within the borders of Switzerland and the Principality of Liechtenstein. It provides vectorial geometries and details such as the Federal Statistical Office identifier number, municipality names, and population. |
| Power distribution grids & load share map (*) | **Source**: Scientific paper and its dataset[12,49] (https://doi.org/10.5281/zenodo.15167589). |
| | **Comment**: *Power distribution grids* - Geo-referenced synthetic medium- and low-voltage distribution grids for Switzerland. *Load share map* - This map provides the percentage of peak electricity demand consumed by commercial and residential buildings. It has a resolution of 100 m $\times$ 100 m across Switzerland |

**Table 1.** Overview of the input data. The symbols △ and * in the left column indicate whether the source is based on measurements or modeled data.

**Battery energy storage systems data.** This work considers BESS co-installation with PV systems. Utility-scale BESSs, primarily deployed at the transmission level[11], are excluded, as the focus is on distribution. Due to the lack of detailed data on small-scale BESS in Switzerland, their allocation, storage capacity, rated power, and charging and discharging efficiency are estimated. Furthermore, because BESS behavior varies significantly with the operational logic, time profiles are not included, as they would not capture the systems' flexibility.

Starting from the PV data, BESS deployment is determined using the co-installation factors listed in Table 2. The 2050 value follows the SFOE projection[30], while the 2030 and 2040 rates are linear interpolations using Europe's BESS/PV co-allocation of 2021[31]. Random subsets of PV systems are selected for the projected years to co-install BESS, matching the deployment rates. All buildings with BESS installations retain them in subsequent years. Moreover, BESS sizing assumes charging and discharging powers equal to the nominal power of the associated PV system. The storage duration and round-trip efficiency are set to 2.5 hours[32] and 85%[33], respectively.

**Heat pumps data.** The building and dwelling register of the Swiss Federal Statistical Office[34] is used to allocate HPs and determine their power rating. This dataset includes geographic coordinates and structural characteristics for residential and commercial buildings in Switzerland. Preprocessing steps are applied to ensure data quality and consistency. Namely, buildings lacking geographic coordinates or having a total heated floor area of less than 5 $m^2$ are excluded, assuming they are not occupied. Missing construction years are filled using the dataset's average value. For buildings without heating area information, estimates are derived by multiplying the total floor area by the number of floors. If the number of floors is unavailable, a single floor is assumed.

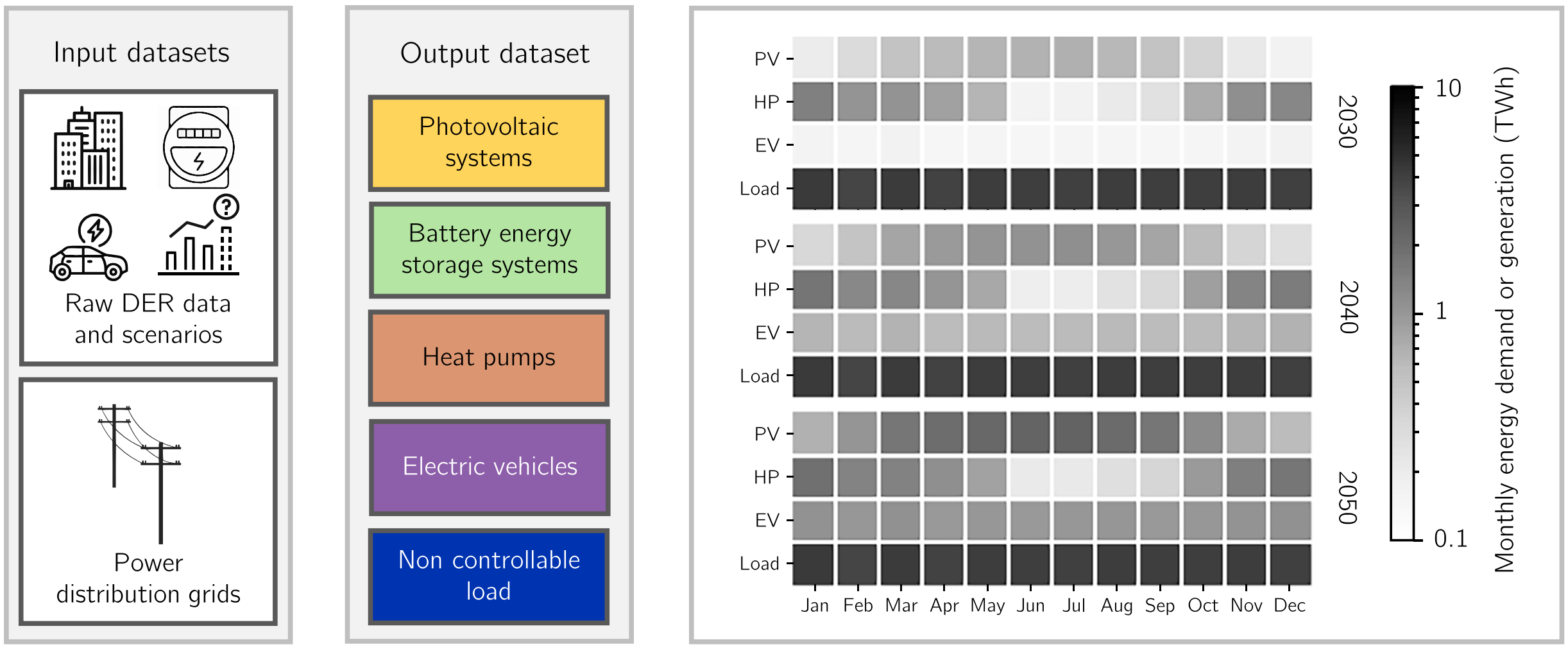

**Fig. 1** Overview of the dataset construction. The input datasets on the left represent the data sources, including raw DERs data, forecast scenarios, and power distribution grids. The output dataset contains information on PVs, BESSs, HPs, EVs, and non-controllable loads integrated into the distribution grids. The right panel presents Switzerland's aggregated monthly electric energy demand and generation from 2030, 2040, and 2050 in the output dataset.

| Resource | Parameter | 2030 | 2040 | 2050 |
|---|---|---|---|---|
| PV | Roof penetration | 27% | 45% | 100% |
| BESS | Co-allocation with PV | 31% | 51% | 70% |
| HP | Commercial penetration | 14% | 21% | 28% |
| | Residential penetration | 36% | 52% | 65% |
| | Fixed COP | 3.49 | 3.79 | 4.12 |
| | $0^{th}$ term of variable COP | 5.60 | 6.08 | 6.61 |
| | $1^{st}$ term of variable COP | $-9.00 \times 10^{-2} K^{-1}$ | $-9.77 \times 10^{-2} K^{-1}$ | $-1.06 \times 10^{-1} K^{-1}$ |
| | $2^{nd}$ term of variable COP | $5.00 \times 10^{-4} K^{-2}$ | $5.43 \times 10^{-4} K^{-2}$ | $5.90 \times 10^{-4} K^{-2}$ |
| | Envelope factor | 90% | 78% | 70% |
| | Temperature | Unchanged | Unchanged | Unchanged |
| EV | Passenger car electrification | 15% | 60% | 100% |
| Demand | Demand | Unchanged | Unchanged | Unchanged |

**Table 2.** Parameters for future scenario projections of distributed energy resources and demand.

After data cleaning, the HPs' thermal parameters for each building are determined based on usage type (commercial, multi-family house, single-family house) and construction period. The thermal transmittance $U_i$ of each building $i \in \mathcal{B}$ is determined. This value is then multiplied by the building's heating area $A_i$ to compute its thermal conductance $H_i$. The thermal transmittance $U_i$ is computed based on statistical analyses of Swiss heating data and a set of building archetypes $\mathcal{K}$, defined by construction period and building use[35]. For each archetype building, the overall thermal transmittance is computed as:

$$U_k = \frac{U_k^r A_k^r + U_k^{wl} A_k^{wl} + U_k^{wd} A_k^{wd} + U_k^f A_k^f}{A^r + A^{wl} + A^{wd} + A^f} \quad \forall k \in \mathcal{K}, \tag{2}$$

where, $U_k^r, U_k^{wl}, U_k^{wd}, U_k^f$ are the thermal transmittance, and $A_k^r, A_k^{wl}, A_k^{wd}, A_k^f$ are the thermal areas of roof, walls, windows, and floor for building archetype $k$, respectively. Then, each building in the Swiss registry[34] is associated with an archetype[35] based on its construction year and use, defining the mapping function $k = K(i)$. Thus, the thermal conductance is computed as:

$$H_i = U_{K(i)} A_i \quad \forall i \in \mathcal{B}. \tag{3}$$

The building archetypes[35] provide specific thermal capacitance values for Swiss buildings, distinguishing between light, medium, and heavy construction types. These are denoted as $c_1$, $c_2$, and $c_3$, equal to 0.1, 0.3, and 0.5 MJ/m$^2$ K. The same study also reports the distribution of construction types across different construction year intervals in Switzerland. Based on these distributions and the construction period of each building, a construction type is randomly assigned, defining the mapping function $l = L(i)$. The building's total capacitance is then obtained by multiplying the specific capacitance by the building's heating area:

$$C_i = c_{L(i)} A_i \quad \forall i \in \mathcal{B}. \tag{4}$$

The nominal HP electrical power installed in each building follows the Swiss sizing guidelines for residential constructions[36]. These guidelines define recommended ranges for specific nominal power depending on the building's construction type. For each type, a representative value is chosen as the midpoint of the recommended range, i.e., $\overline{p_1^{HP}} = 60\,\text{W/m}^2$, $\overline{p_2^{HP}} = 45\,\text{W/m}^2$, and $\overline{p_3^{HP}} = 35\,\text{W/m}^2$ for heavy, medium, and light constructions, respectively. The nominal HP power $\overline{p_i^{HP}}$ assigned to each building is computed by multiplying the specific nominal power associated with its construction type $L(i)$ by its heating area $A_i$:

$$\overline{P_i^{HP}} = \overline{p_{L(i)}^{HP}} A_i \quad \forall i \in \mathcal{B}. \tag{5}$$

Each building is linked to an hourly outdoor temperature profile based on the closest of 441 weather stations maintained by MeteoSwiss[37], using 2023 hourly temperature data. The power consumption of a given HP installation is then modeled as:

$$\beta_{i,t} P_{i,t}^{HP} = C_i (T_{i,t} - T_{i,t-1}) + H_i (T_{i,t} - T_{s(i),t}) \quad \forall i,\, t, \tag{6}$$

$$0 \leq P_{i,t}^{HP} \leq \overline{P_i^{HP}} \quad \forall i,\, t, \tag{7}$$

where the subindex $i$ refers to the $i$-th building, and $t$ denotes time. $\beta_{i,t}$ is the coefficient of performance (COP), $P_{i,t}^{HP}$ is the HP power consumption, $T_{i,t}$ is the indoor temperature, $T_{s(i),t}$ is the temperature at the weather station $s(i)$ corresponding to the building, and $\overline{p_i^{HP}}$ is the nominal installed HP power. This formulation considers that HPs deliver heating power only, excluding cooling due to Switzerland's mild maximum temperatures.

The DER dataset provides the COP $\beta_{i,t}$ in two forms: fixed and temperature dependent. Users can select the appropriate form based on application requirements. In the fixed formulation, the COP of an HP remains constant across all operational time-steps. In the variable formulation, the COP follows a quadratic relation:

$$\beta_{i,t} = \alpha_0 + \alpha_1 \Delta_{i,t} + \alpha_2 \Delta_{i,t}^2 \quad \forall i,\, t, \tag{8}$$

where $\Delta_{i,t} = T_{i,t} - T_{s(i),t}$, and $\alpha_0$, $\alpha_1$, and $\alpha_2$ are the 0$^{\text{th}}$-, 1$^{\text{st}}$-, and 2$^{\text{nd}}$-order coefficients, respectively.

Finally, HP data is generated for 2030, 2040, and 2050, reflecting Switzerland's device penetration, insulation, and performance projections[30], as shown in Table 2. First, buildings are uniformly randomly selected to match HP penetration rate projections for the commercial and residential sectors. To preserve temporal consistency, buildings selected in a given year retain their HP installations in subsequent years. Conductance values are scaled using envelope factors of 90%, 78%, and 70% in 2030, 2040, and 2050 to account for improvements in building insulation. Furthermore, the fixed COP of each HP is set according to the projections for each respective year. Variable COP coefficients for 2030 follow current literature values for air-to-water HPs[38], while the coefficients for 2040 and 2050 are derived by scaling those of 2030 with the expected improvements[30].

**Electric vehicles data.** The electric vehicles dataset used in this work is derived from a mobility dataset covering over 3.2 million synthetic EVs across Switzerland[39]. It includes geo-referenced trips, parking patterns, and driving energy demands, calculated using a detailed traction, heating, cooling, and regenerative power model[40]. Given the volume and granularity of the original data[41,42], it is incorporated here in an aggregated form at the municipality level. For each municipality, four components are provided: the hourly uncontrolled EV base charging profile, upper and lower bounds for flexible charging activation, and the daily flexible energy, which indicates the total energy that can be shifted while respecting mobility constraints. The power profiles are given with an hourly resolution over a year, while the flexible energy profiles have a daily resolution. Importantly, when controlling flexible charging, the weekly energy consumption must match the base charging energy to preserve the estimated total charging demand. The input dataset corresponds to the year 2050, assuming total EV penetration. To generate profiles for 2030 and 2040, the 2050 power and energy values are scaled by the projected EV penetration levels[30] reported in Table 2.

**Non-controllable load data.** In this context, non-controllable loads refer to conventional electricity consumption from residential, commercial, and industrial users, excluding EVs and HPs. Max-normalized (i.e., divided by the maximum value) yearly non-controllable load profiles are obtained for the nodes in low- and medium-voltage grids. As later explained, these normalized profiles are then scaled by the peak non-controllable active power load at each node to derive the corresponding nodal profiles.

Low-voltage non-controllable demand consists mainly of residential and commercial electricity consumption. These demand types are modeled using municipality-resolved residential and commercial load profiles, derived by aggregating the representative Swiss building demand time series[43,44]. The profiles represent yearly electricity demand with hourly resolution for building archetypes defined by usage type and construction period[34]. The number of each archetype in every municipality is counted and multiplied by its respective profile.

This yields cumulative residential and commercial demand profiles per municipality. These profiles are then normalized by their peak power to obtain hourly max-normalized commercial and residential time series for each municipality.

For medium-voltage grids, load profiles corresponding to industrial and large commercial consumers are assigned to the medium-voltage nodes. For this purpose, only a single Swiss medium-voltage load profile obtained for Zurich[45] is publicly available, excluding contributions from the low-voltage level. The dataset spans the years 2015 to 2025 and has a 15-minute resolution. An hourly load curve is computed by aggregating the data to hourly resolution, averaging across years, and normalizing the resulting profile by the maximum value.

Non-controllable load profiles are assumed to remain unchanged for 2030, 2040, and 2050. This invariance reflects the assumption that their behavior will remain constant across these future years.

**Grid integration.** The synthetic distribution grids for Switzerland used in this work comprise 879 medium-voltage and 34,920 low-voltage grids, operating at 20 kV and 400 V, respectively[12]. These models include specifications for power lines and transformers. Additionally, all lines and nodes are geo-referenced, and the nodes include the estimated local peak load. Table S1 of the Supplementary Information summarizes grid characteristics by canton and for the country.

The identified DER information is integrated into the distribution grids by assigning resources to the nodes. It includes the locations and parameters of PVs, BESSs, and HPs at building resolution. For EVs, the data consists of projected municipal-level uncontrolled base charging profiles, upper and lower bounds for flexible charging activation, and daily flexible energy. In addition, normalized non-controllable load profiles represent the electricity consumption of residential and commercial buildings in each municipality and medium-voltage consumers across Switzerland.

Rooftop PVs and co-located BESSs primarily connect to the low-voltage level, except for larger systems that connect to the medium-voltage level, as detailed below. Since BESSs are co-located with PVs, their placement follows the PV allocation logic. The distribution of rooftop PV system power ratings is long-tailed, so statistical cutoffs are used to segment devices by voltage level. A standard one-sided statistical trimming method[46] selects the top 5% of values and assigns large PV systems to the medium-voltage level to avoid unrealistic power rating conditions. The remaining 95% connect to their nearest low-voltage nodes. Maximum distance thresholds are imposed to address the mismatch between the buildings' dataset used for the distribution grid models and the rooftop PVs dataset. The distribution grid models[12] are based on buildings from the OpenStreetMap database[47], while the PV input dataset considers the Swiss building registry[48].

Devices located more than 50 m from any low-voltage node are excluded. PV systems assigned to the medium-voltage grid connect to their closest medium-voltage node, excluding devices located more than 3 km from any node. A further 5% trimming is applied to the medium-voltage installations to exclude massive systems, which are assumed to be connected at higher voltage levels and are, therefore, not included in the dataset. A cutoff is applied at the nodal level to prevent excessive PV allocation to single nodes and thus reduce cases of extreme feeder loading. Hence, the nodal PV peak power is capped at 100 kW for low-voltage nodes[25] and 1 MW for medium-voltage nodes[26].

The placement of HPs follows a similar approach. After identifying all HPs, a 5% upper-tail cutoff filters out larger units and allocates them to the medium-voltage level. The remaining 95% connect to their nearest low-voltage nodes. Again, devices more than 50 m away from a node are neglected. HPs associated with the medium-voltage level connect to their nearest node, excluding those more than 3 km away. A further 5% trimming is applied to the medium-voltage installations to exclude the very large systems. Furthermore, the same nodal peak power limits used for PVs are applied to HPs.

EVs can be charged at various locations. Therefore, a simplified approach is adopted to distribute the aggregated municipal EV charging power and flexible energy profiles. These profiles are proportionally allocated to the low-voltage nodes according to each node's share of the municipality's non-controllable consumption peak.

The normalized commercial and residential profiles per municipality are combined at the low-voltage nodal level using weighted averages based on the nodal commercial and residential weights[49]. The nodal consumption is then derived by multiplying each node's peak non-controllable demand[49] by its corresponding weighted average profile.

Medium-voltage non-controllable load data is assigned to the final medium-voltage consumption nodes (i.e., nodes that are not medium-to-low-voltage substations) using the normalized profile. As with low-voltage loads, multiplying the normalized profile by the nodal peak power consumption[49] yields the medium-voltage non-controllable demand.

The integration of DERs and non-controllable load data into the distribution grids is exemplified in Fig. 2. The right panel shows all distribution grids in Switzerland, where dark green and light orange lines represent the medium- and low-voltage grids, respectively. The left panel zooms into one low-voltage grid, with pie charts at the nodes indicating the magnitude and share of DERs and non-controllable load for the year 2050. Hence, each grid contains DERs and non-controllable load information at nodal resolution. The nodal power allocation distributions in Switzerland for PVs, BESSs, and HPs at low- and medium-voltage levels appear in Figure S1 of the Supplementary Information.

For illustration purposes, HP consumption profiles are obtained by applying a fixed-temperature thermostat that maintains a constant indoor temperature of 21° C, using Eqs. (6)-(7) with fixed COPs. Nodal-resolved HP consumption profiles are omitted in the provided dataset, as they would require billions of entries. They are generated using the function `compute_hp_consumption` in the Python script Data_checker.py provided with the dataset. Figure 3 presents aggregated PV generation, HP demand, EV base charging, and non-controllable demand for the same sample low-voltage grid shown in Fig. 2 for 2030, 2040, and 2050. As the penetration of PVs, HPs, and EVs increases, the magnitude of their base profiles grows accordingly. Moreover, PV and HP behavior show clear daily and seasonal patterns: PV generation peaks around noon, while HP consumption reaches its maximum at night. Similarly, monthly PV generation is highest in summer, whereas HP demand peaks in winter. These patterns also appear in the aggregated profiles across all distribution grids, as shown in Figure S2 of the Supplementary Information.

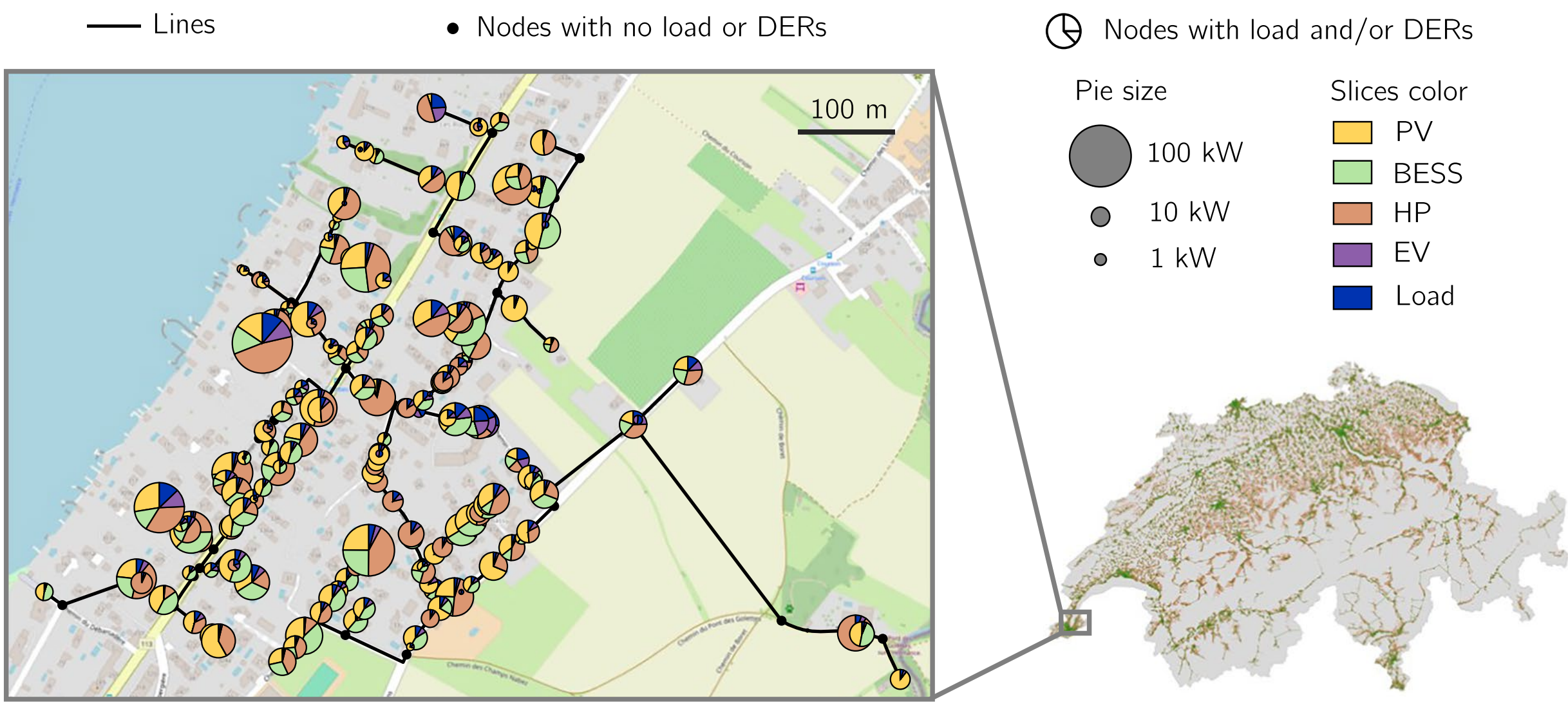

**Fig. 2** The bottom right shows the distribution grid models for Switzerland, with the gray background representing the country's political boundaries. Dark green and light orange lines indicate the medium- and low-voltage grids, respectively. The top right presents the legend for the plot on the left, which provides a closer view of the low-voltage grid 6602-2_0_5. The grid nodes illustrate the magnitude and share of DERs and non-controllable load in 2050. PV power refers to the installed power rating, co-located BESS power to the charging/discharging power, HP power to electrical power, EV power to the peak of the base charging profile, and likewise to the non-controllable load profile.

**Operational flexibility parametrization.** The generated dataset enables users to model DER flexibility capabilities at the nodal level, illustrated in Fig. 4 for a selected low-voltage node on a day in 2050. Nodal PV installed capacity and active power generation profiles are provided, allowing for the representation of power curtailment. Users can model the SOE of BESSs, considering charging and discharging efficiencies, nominal power, and energy storage capacity at the nodes. HP heating is controllable to maintain the indoor temperature within a thermal comfort band defined by the modelers. For this purpose, the dataset incorporates building thermal capacitance and conductance at the nodes, HP coefficients of performance, device power ratings, and outdoor temperature profiles. Lastly, EV charging can be controlled for every node. This control requires that charging remains within the upper and lower flexible charging activation profiles. In this regard, the daily sum of absolute deviation from the base profile must be equal to or smaller than the daily flexible energy profile. Moreover, the controlled EV charging energy during a week must be the same as the energy consumed by the base charging profile.

In this context, the variability of daily and seasonal DER behavior, as shown in Fig. 3, significantly influences the temporal availability of flexibility, which the dataset also captures. Since PV generation occurs during daylight hours, curtailment capabilities are similarly limited to that period. BESSs provide charging and discharging capabilities, enabling generation transfer to meet consumption. As outdoor temperatures tend to drop at night, HPs must operate to prevent inadmissibly low indoor temperatures, making night-time heating shifts relevant for flexibility. EVs exhibit lower base charging in the early morning, increasing the potential for shifting to anticipate charging needs later in the day. Seasonal variation also affects the relevance of DERs for flexibility: HP flexibility becomes more important in winter due to a higher share of energy consumption, while PV generation increases in summer, making PV-based flexibility more applicable during that season. Therefore, the differences in the temporal capabilities of DERs highlight the importance of modeling different technologies to diversify flexibility sources.

## Data Record

The dataset is publicly available on Zenodo (https://doi.org/10.5281/zenodo.15056134)[50]. It includes 57 comma-separated values (CSV) files, with 19 files for each projected year (2030, 2040, and 2050). The CSV files contain data on PV installations, PV generation profiles, co-located BESS installations, HP installations, thermal descriptions of served buildings, outdoor temperature time series, EV profiles, and non-controllable load profiles. Power, temperature, and energy values are expressed in kW, $^\circ$C, and kWh. Unique identifiers link devices and time profiles to distribution grid nodes. For simplicity, each year has 365 days and begins on a Monday at 00:00. February 29th of 2040 is omitted. Time series data have standardized resolutions of one hour for power and temperature profiles and one day for EV daily flexible energy profiles. A folder containing distribution grid data is provided, in addition to a complementary folder including municipality boundaries and a Python script for data loading. Tables 3 and 4 summarize the dataset's structure and content, further detailed in Table S2 of the Supplementary Information.

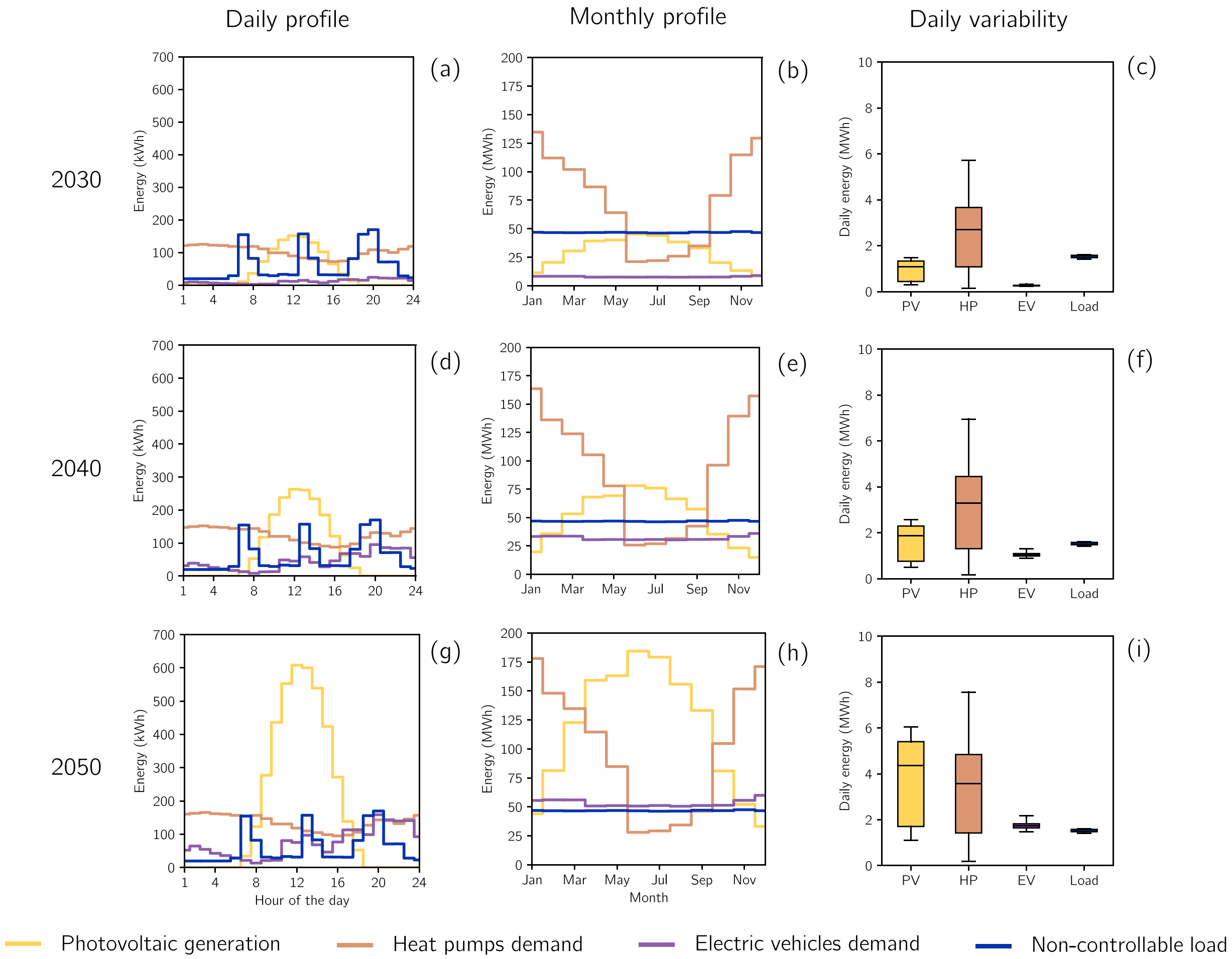

**Fig. 3** Daily profiles, monthly profiles, and daily variability graphs are shown for the low-voltage grid 6602-2_0_5. Plots (**a**), (**b**), and (**c**) correspond to projections for 2030; (**d**), (**e**), and (**f**) to 2040; and (**g**), (**h**), and (**i**) to 2050.

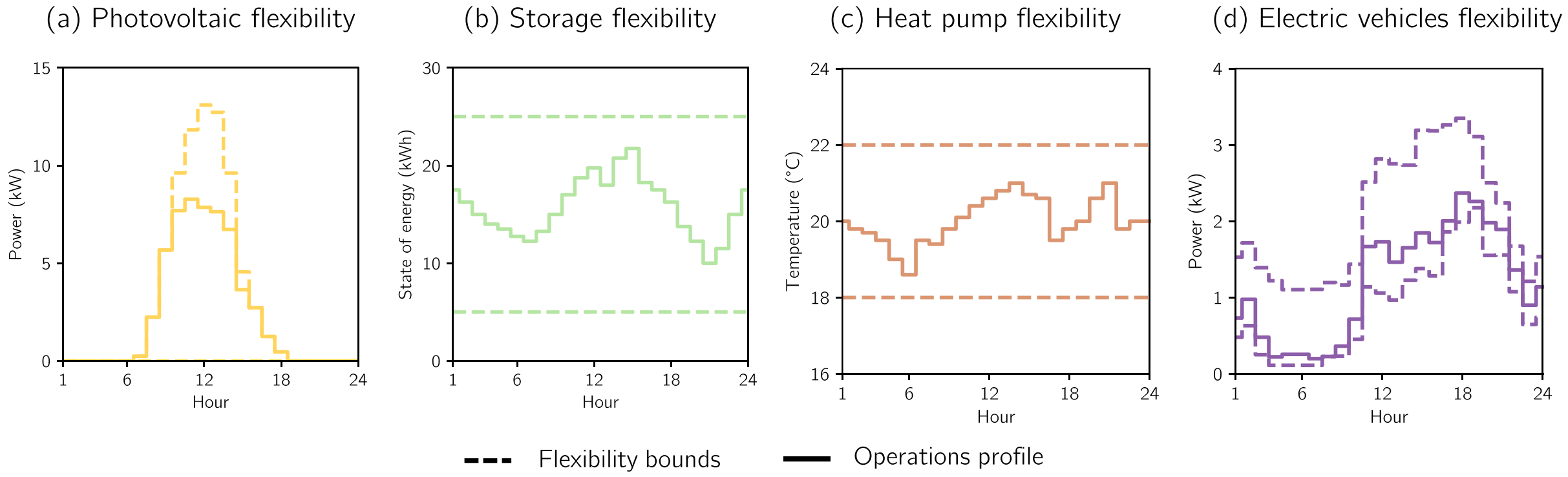

**Fig. 4** Flexibility provision capabilities modeled using the database for node 147 of the low-voltage grid 6629-4_0_4 for March 31st of 2050. Plot (**a**) shows a dashed line representing the maximum available active power generation over time, while the continuous line depicts a possible profile with curtailment. Plot (**b**) presents a schematic of a BESS's SOE, with a continuous line indicating the SOE and dashed lines representing upper and lower energy storage limits. Plot (**c**) illustrates the indoor temperature of a building, which HPs can control, along with upper and lower comfort temperature bounds. Plot (**d**) represents the base EV charging profile with a continuous line and upper/lower charging shifting profiles with dashed lines.

| Output files | | |
|---|---|---|
| **Archive folder** | **Year folder** | **Files** |
| 01_PV | 2030, 2040, 2050 | *LV_generation.csv* - PV active power generation profiles for low-voltage nodes (kW). |
| | | *LV_P_installed.csv* - Nominal installed PV capacity in kWp for the low-voltage nodes. |
| | | *LV_std.csv* - Standard deviation for PV active power generation profiles for low-voltage nodes (kW). |
| | | *MV_generation.csv* - PV active power generation profiles for medium-voltage nodes (kW). |
| | | *MV_P_installed.csv* - Nominal installed PV capacity in kWp for medium-voltage nodes. |
| | | *MV_std.csv* - Standard deviation PV active power generation profiles for medium-voltage nodes (kW). |
| 02_BESS | 2030, 2040, 2050 | *BESS_allocation_LV.csv* - The parameters of the BESS installed at the low-voltage nodes. |
| | | *BESS_allocation_MV.csv* - The parameters of the BESS installed at the medium-voltage nodes. |
| 03_HP | 2030, 2040, 2050 | *LV_heat_pump_allocation.csv* - Nominal installed electrical HP capacity in kW for low-voltage nodes, the thermal capacitance/conductivity of the served buildings per node, the HPs coefficient of performance (fixed and variable), and the outdoor temperature profile name identifier. |
| | | *MV_heat_pump_allocation.csv* - Nominal installed electrical HP capacity in kW for medium-voltage nodes, the thermal capacitance/conductivity of the served buildings per node, the HPs coefficient of performance (fixed and variable), and the outdoor temperature profile name identifier. |
| | | *Temperature_profiles.csv* - Yearly, hourly resolved, ambient temperature profile (in degrees Celsius) from weather stations. |
| 04_EV | 2030, 2040, 2050 | *EV_power_profiles_LV.csv* - The base profile represents the uncontrolled charging demand, while the upper and lower bounds define the maximum and minimum power levels, respectively (kW). |
| | | *EV_flexible_energy_profiles_LV.csv* - Maximum flexible energy that can be shifted per day from the base power charging profile (kWh). |
| | | *EV_allocation_LV.csv* - Share of municipality-level EV profiles assigned to low-voltage nodes. |
| 05_Demand | 2030, 2040, 2050 | *LV_basicload_shares.csv* - Share of commercial and residential demand for each low-voltage node. |
| | | *Commercial_profiles.csv* - Hourly max-normalized load profile for commercial demand at the municipality level, identified by the BFS municipality code. |
| | | *Residential_profiles.csv* - Hourly max-normalized load profile for residential demand at the municipality level, identified by the BFS municipality code. |
| | | *MV_load_profile.csv* - Representative hourly max-normalized non-controllable load profile for medium-voltage nodes. |

**Table 3.** Data repository and its description.

## Technical Validation

The dataset is validated at three levels. First, the integrity and logical consistency of the data are verified. Second, the coherence of energy consumption and generation is analyzed. Lastly, the geographical distribution of DERs is evaluated against the population distribution.

**Integrity and logical consistency of the data.** The integrity and logical consistency of the output files are verified using automated code (available with the dataset). These procedures ensure that the output files meet several necessary criteria. First, the output files contain the expected number of columns and rows, as detailed in Table S2 of the Supplementary Information. Second, the column data types are validated, confirming that numeric and categorical values match the expected type. Third, completeness is verified, ensuring no empty entries appear in the files. Lastly, consistency checks assess relationships between variables: upper EV charging power values need to be equal or greater than the corresponding base charging profiles, which in turn need to be equal or greater than the lower EV charging power values; daily flexible energy cannot exceed the sum of the maximum absolute deviations between the base and upper/lower profiles; and PV installed capacity always needs to be equal to or greater than the available generation.

**Coherence of energy consumption and generation.** The coherence of energy consumption and generation in the dataset is assessed by comparing the results with the Energy Perspectives 2050+ report from the SFOE[30]. This report presents scenario-based analyses exploring possible technological developments through which the country can simultaneously achieve energy and climate policy goals. Table 5 reports the dataset's national values, consumption projections from Energy Perspectives[30], and PV rooftop generation[27] scaled by the expected deployment reported in Table 2.

First, the dataset assigns massive PV installations to voltage levels above those of distribution grids and omits geographically distant buildings. As a result, the PV generation allocated to distribution grids is lower than that of the complete building dataset. Specifically, a 6.4 TWh discrepancy exists in the 2050 PV generation. Three factors explain this: 1) 1.7 TWh from buildings are excluded due to excessive distance from distribution nodes; 2) 2.7 TWh from large installations are assumed to connect at higher voltage levels; and 3) 2.0 TWh are left out due to nodal-level capping of PV capacity. Second, the HP demand in the dataset reflects the inclusion of large HPs connected to medium-voltage grids, which the report does not include in the building HP category. Energy Perspectives aggregates the consumption of large HPs, electrolyzers, and carbon capture and storage systems into a single category, which prevents isolating the HP contribution. Third, the dataset's EV charging energy consumption aligns with the report's values, which are slightly higher for every modeled year. Finally, the report's projections for final non-controllable electricity consumption closely match the dataset in 2030 but decline in later years due to expected energy efficiency measures.

| Additional folders | |
|---|---|
| **Archive folder** | **Description** |
| 06_Grids | In this repository, only the grids for the integrated medium-low voltage system 459_0 are provided. To access all available grids, refer to the original grid repository[49] and replace the LV and MV zip files accordingly. For further details on the repository, refer to[12]. The folder contains the following files: |
| | *LV.zip:* All the low-voltage power distribution grids data, such as grid topology, branch flow limits, line impedance, and nodal peak powers. |
| | *MV.zip:* All the medium-voltage power distribution grids data, such as grid topology, branch flow limits, line impedance, and nodal peak powers. |
| | *dict_folder.json:* A dictionary that maps the BFS municipality number, the number before the dash (-) of the low-voltage grid code, to the corresponding subfolder in the LV.zip archive. This file is used to access the low-voltage grid data. |
| 07_Complementary_data | This folder contains complementary data, which are not needed to use the DERs dataset but may be of use for its interpretation and modification. It contains the following file: |
| | *Municipalities_2022_01_crs2056.geojson:* This GeoJSON file contains the municipal boundaries of Switzerland as part of the swissBOUNDARIES 3D dataset[51]. The dataset includes detailed geometric representations of administrative units in Switzerland, the Principality of Liechtenstein, and border exclaves of neighboring countries. The municipalities are represented as polygons, based on EPSG:2056 coordinate reference system, and corresponds to the municipalities in 2022. |
| **Additional files** | |
| File | Description |
| 08_Data_loader.py | This Python script provides an example of how to load distributed energy resources (DERs) data for low-voltage and medium-voltage grids for a specified simulation year and time interval. The script is designed to facilitate the loading and processing of the DER data present in the dataset. |

**Table 4.** Additional data included in the repository.

| | 2030 | | 2040 | | 2050 | |
|---|---|---|---|---|---|---|
| **Resource** | **Dataset** | **Reference** | **Dataset** | **Reference** | **Dataset** | **Reference** |
| PV | 5.4 | 6.6 | 8.7 | 11.1 | 18.2 | 24.6 |
| HP | 8.8 | 5.6 | 10.7 | 7.5 | 11.6 | 8.7 |
| EV | 1.8 | 2.4 | 7.2 | 8.1 | 12.0 | 13.1 |
| Load | 48.8 | 49.9 | 48.8 | 45.9 | 48.8 | 41.4 |

**Table 5.** Energy consumption and generation by resource, in TWh, for 2030, 2040, and 2050 in the generated dataset and in reference sources.

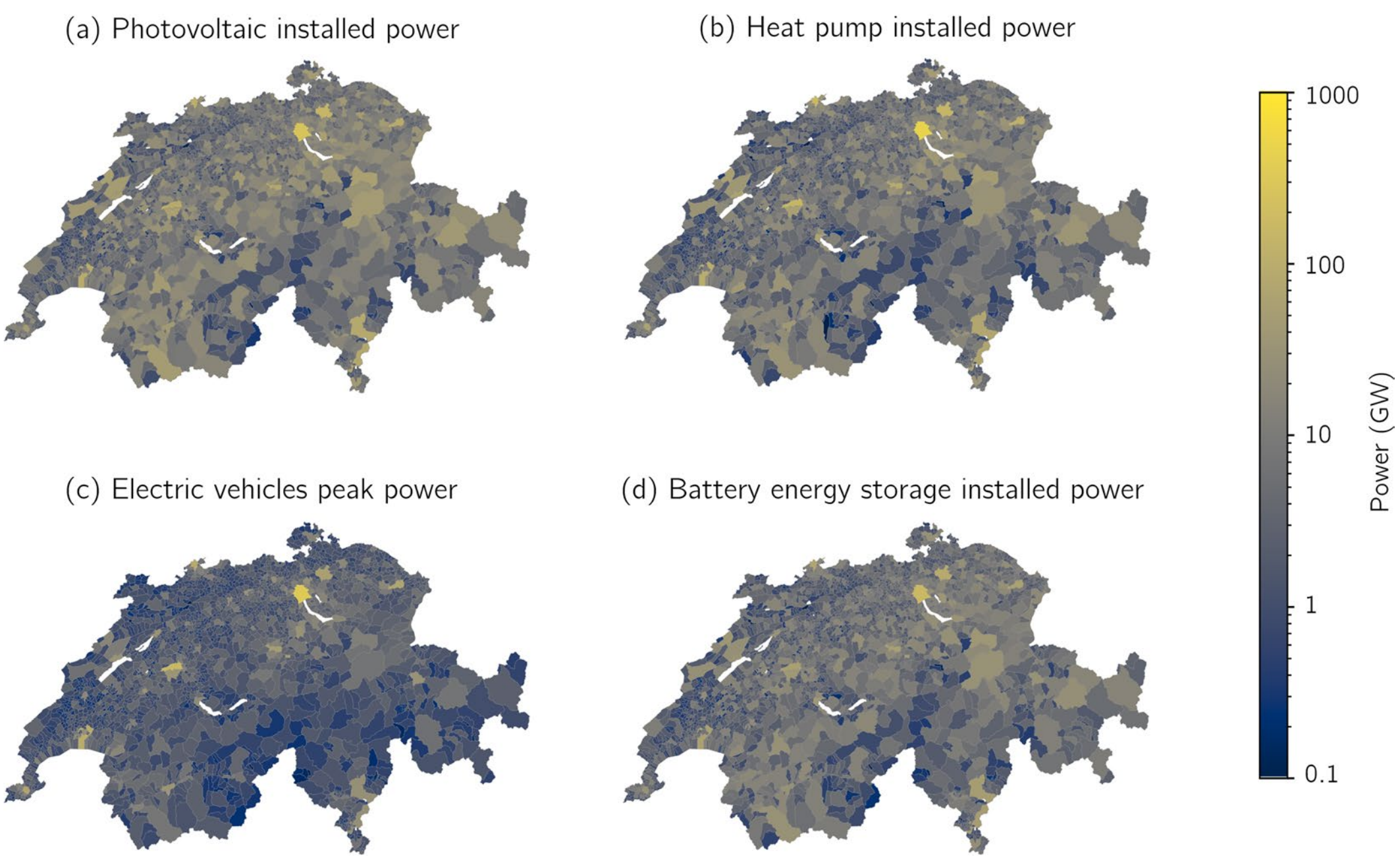

**Fig. 5** The plots show aggregated power values for Swiss municipalities projected to 2050. Plot (**a**) displays installed PV power, (**b**) installed electrical HP power, (**c**) peak EV base profile charging power, and (**d**) installed BESS charging/discharging power.

| Correlations with population distribution | | |
|---|---|---|
| Distributed energy resource | Spearman correlation | Pearson correlation |
| PV | 0.76 | 0.65 |
| BESS | 0.90 | 0.83 |
| HP | 0.93 | 0.93 |
| EV | 0.96 | 0.95 |

**Table 6.** Spearman and Pearson correlation coefficients between the power allocation of distributed energy resources and the population per municipality are presented.

**Geographical allocation of distributed energy resources.** The geographical allocation of DERs is evaluated across Switzerland. Figure 5 presents the aggregated power values for municipalities projected for 2050: Plot (a) shows installed PV power, (b) installed electrical HP power, (c) peak EV base charging power, and (d) allocated BESS charging/discharging power. As expected, these values correlate with municipal population, with higher power levels observed in more populated areas. The Spearman and Pearson correlation coefficients between the aggregated power values and the current municipal population[51] confirm this trend, as shown in Table 6. These relationships are illustrated in Figure S3 of the Supplementary Information, which includes a plot of the aggregated power values against municipal populations.

## Usage Notes

All DER and non-controllable load data are stored as CSV files in a human-readable format. The dataset repository includes detailed instructions for retrieving time series and DER allocations for specific distribution grids. Furthermore, a Python script for data loading is provided with the dataset to simplify its use.

## Code availability

The code used for the dataset generation, analysis, and visualization was implemented in Python 3.11 and is available on GitHub (https://github.com/lorenzozapparoli/SwissDN-DERs). The repository includes a README.md file with documentation, and the code is well-commented.

## Acknowledgements
The research published in this publication was carried out with the support of the Swiss Federal Office of Energy SFOE as part of the SWEET EDGE project. The authors bear sole responsibility for the conclusions and the results. The authors would like to thank Ruilong Wang for her assistance with data retrieval and processing, and Anna Varbella for her suggestions on data and code sharing. Also, the authors acknowledge the valuable inputs that Philipp Schütz and Selina Kerscher gave during the preparation of the manuscript.

## Author contributions

L.Z.- Data processing, analysis, visualization, code development, and original manuscript draft. A.O.- Data processing, analysis, visualization, code development, and original manuscript draft. M.P.H.- Data processing and manuscript editing. B.G.- Manuscript review and supervision. G.H.- Manuscript review and supervision. G.S.- Manuscript review and supervision.

## Funding

Open access funding provided by Swiss Federal Institute of Technology Zurich.

## Competing interests

The authors declare no competing interests.

## Additional information

**Supplementary information** The online version contains supplementary material available at https://doi.org/10.1038/s41597-025-05830-y.

**Correspondence** and requests for materials should be addressed to G.S.

**Reprints and permissions information** is available at www.nature.com/reprints.

**Publisher's note** Springer Nature remains neutral with regard to jurisdictional claims in published maps and institutional affiliations.